# Capture Agent Free Biosensing using Porous Silicon Arrays and Machine Learning

*Simon J. Ward, Tengfei Cao, Xiang Zhou, Catie Chang, Sharon M. Weiss\**

S. J. Ward, C. Chang, S. M. Weiss
Department of Electrical and Computer Engineering, Vanderbilt University, Nashville, Tennessee 37235, USA
sharon.weiss@vanderbilt.edu

T. Cao
Department of Interdisciplinary Material Science, Vanderbilt University, Nashville, Tennessee 37235, USA

X. Zhou
Department of Chemistry, Vanderbilt University, Nashville, Tennessee 37235, USA

Biosensors are an essential tool for medical diagnostics, environmental monitoring and food safety. Typically, biosensors are designed to detect specific analytes through functionalization with the appropriate capture agents. However, the use of capture agents limits the number of analytes that can be simultaneously detected and reduces the robustness of the biosensor. In this work, we report a versatile, capture agent free biosensor platform based on an array of porous silicon (PSi) thin films, which has the potential to robustly detect a wide variety of analytes based on their physical and chemical properties in the nanoscale porous media. The ability of this system to reproducibly classify, quantify, and discriminate three proteins is demonstrated to concentrations down to at least 0.02g/L (between 300nM and 450nM) by utilizing PSi array elements with a unique combination of pore size and buffer pH, employing linear discriminant analysis for dimensionality reduction, and using support vector machines as a classifier. This approach represents a significant step towards a low cost, simple and robust biosensor platform that is able to detect a vast range of biomolecules.

# 1. Introduction

Biosensors are devices used to detect biological analytes for applications including medical diagnostics, environmental monitoring and food safety. Typically, a biosensor is composed of two primary components: a capture agent (also sometimes referred to as a probe molecule or bioreceptor), which specifically binds to the desired target analyte, and a transducer, which converts the binding of the target species to a measurable optical, colorimetric, electrochemical, thermal, or microelectromechanical signal.[1] Examples of specific capture agent-target analyte 'lock and key' interactions are antibody-antigen interactions, enzyme-substrate interactions, peptide interactions, and oligonucleotide interactions.[2] However, the reliance on capture agents for analyte detection brings several disadvantages. The first is that a capture agent, by design, typically only binds to one species. If the aim is to identify and quantify multiple molecular constituents in a solution of unknown composition, multiple capture agents are required, which leads to larger, more complex and expensive biosensors. Furthermore, if there is insufficient capture agent coverage for all molecules which may be present, the possibility of dangerous species going undetected can be a concern. In addition, many capture agents denature or degrade over time and/or in harsh environments, which limits shelf life, ease of transportation and the types of locations where the sensor can be used. Finally, when challenged with detecting a new target molecule of interest for which there is no existing capture agent, it can take significant time to develop an effective capture agent with sufficient specificity and affinity.

Fourier transform infrared spectroscopy and Raman spectroscopy are common tools that have been employed to identify molecular species without the use of a capture agent; however, these approaches require costly and generally bulky instrumentation, making them unsuitable for many applications. In addition, there are significant challenges associated with identifying molecules in a complex background using these approaches, including performing measurements in the aqueous phase.[3]

To address these issues, sensor arrays have been developed to mimic the mammalian olfactory system, which is able to detect over 10,000 different volatile odorants by using hundreds of cross-reactive olfactory receptor epithelial cells with varying affinity for a wide range of molecules and pattern recognition in the brain.[4,5] In synthetic 'electronic noses',[6–8] the collection of olfactory receptors are replaced with array-based sensors relying on one of the following mechanisms: optical sensors comprising fluorometric or colorimetric dyes,[9] electrochemical sensors such as amperometric[10] and metal oxide semiconductor sensors,[11] and microelectromechanical mass sensors such as piezoelectric[12] and surface acoustic wave (SAW) sensors.[13] Non-specific receptors in previously reported sensing arrays can be divided into two categories, either relying on chemical interactions or physical interactions with the target analyte.

The primary advantage of utilizing chemical interactions as the basis of a sensor array is the relative strength of these interactions, which can be orders of magnitude higher than physical interactions. However, careful consideration must be given to the design of such an array in order to ensure that sufficient chemical reactivity space is probed to distinguish any molecule which may be encountered. Furthermore, the mechanical and chemical stability and robustness of sensing arrays based on chemical interactions can be an issue, and complexity of design and manufacture is often high. On the other hand, sensor arrays employing physical interactions are usually robust and stable, but they can suffer from poor sensitivity and low dimensional discriminatory space due to the relatively weak interactions and similarity of response between molecules. Additionally, different sensor parameters are often highly correlated for sensor arrays relying on physical interactions, resulting in the majority of the variance (>95%) and discriminatory ability being captured in one dimension: a linear combination broadly representing the property of hydrophobicity.[6]

We also note that different sensor platforms used in cross-reactive sensor arrays have their own inherent limitations. For example, colorimetric sensor arrays which have been the subject of wide research interest, are limited in: the number of array elements, due to the complexity of manufacture, reproducibility, and printing quality; the minimum spot size of chemo-responsive dyes, as a result of edge effects and limited printing resolution; and their use in the aqueous phase which often results in underwhelming detection limits (>1μM).[14] Finally, many specific capture agent-free biosensors rely on bioreceptors designed to be cross-reactive, including enzymes, antibodies, and aptamers,[8] which introduce many of the same limitations as specific capture agents – changing response and structure and potentially denaturing in harsh environments, poor shelf life, and expensive and complex manufacturing process.

In this work, we show proof-of-concept results demonstrating that an array of porous silicon (PSi) sensors can be used to classify, quantify and discriminate three proteins – bovine serum albumin (BSA), chicken ovalbumin, and avidin – at concentrations down to 300nM with 100% accuracy when the discrete set of possible concentrations is known, and 87.5% for unseen concentrations, without the use of capture agents. The proteins, suspended in buffer solutions with varying pH, are exposed to PSi films with different pore sizes. The proteins were chosen based on their distinct combinations of isoelectric points and molecular weights, and because they are well characterized which allows for incorporation of prior knowledge into the machine learning analysis. However, this approach is generalizable to any molecule. This sensor array relies on physical rather than chemical adsorption, enabling straightforward classification and quantification of biomolecules in the liquid phase. Specifically, molecules which diffuse into the pores adsorb to the pore walls via electrostatic, dehydration, hydrogen bonding, dipole-dipole and Van der Waals interactions, which are governed by a molecules net charge and charge distribution, hydrophobicity, and polarity amongst other characteristics.[15] These properties will provide some limited discrimination between molecules; however, the response of similar molecules is approximately homogenous, hence the need to introduce other variables with a much larger differential response, such as pore size and environment pH. The inherent vast surface area of PSi helps circumvent the issue of weak physical interactions by presenting many more adsorption sites and enhancing interaction between light and molecules adsorbed in the pores. Furthermore, the parameter space is not limited solely to hydrophobicity as is the case with other sensors relying on physisorption, since pH and pores size distribution also independently influence adsorption profiles of molecules according to isoelectric point (pI), and molecular size and shape, respectively.

## 2. Results and Discussion

Solutions of different concentrations of BSA, chicken ovalbumin, and avidin were exposed to PSi sensor arrays with six distinct elements, each having a unique combination of solution pH and PSi pore size. Sixteen of these sensor arrays were constructed for every concentration of each protein, containing six elements (3 pore sizes and 2 pH values), and randomly sampled pairs of array measurements were averaged to reduce variance of response, yielding 8 independent repeats. Linear discriminant analysis (LDA) was used to reduce dimensionality of the sensing array measurements, originally 6D, down to 3D, to visualize the degree to which the proteins are separable, and support vector machines (SVM) were used to classify protein response matrices.

PSi thin films were fabricated using three different current densities, achieving three distinct pore size distributions with a different average pore size. The characterization of each of the three films is shown in Figure 1, which illustrates the distribution of pore diameters for each etching current density used, as obtained through SEM image analysis, and the corresponding baseline reflectance spectra for the PSi films. The pore size distributions are weighted by the circumference of the pores, which is proportional to the number of binding sites. Table 1 shows the average pore diameter and thickness, measured by analyzing SEM images, and porosity, calculated from optical reflectance measurements, as a function of etching current density. The thickness of the PSi films was held approximately constant for all samples through the choice of the etching time. Both Figure 1 and Table 1 show the pore size distributions shifting to larger diameters as the etching current density is increased. Importantly for size selectivity, the fraction of larger pores dramatically increases by an order of magnitude with the etching current density. We believe this metric is a dominant effect governing the response of the PSi films to protein exposure compared to the average pore size, which exhibits only a modest increase with etching current density. We note that our analysis of top view SEM images is only intended to provide a rough estimate of the pore size and establish the general trend of increasing pore size with increasing etching current density. Thresholding to convert the images from greyscale to binary was done manually and the fine structure of the pore branches within the pores was not rigorously taken into account. Solutions of the three proteins, at three different concentrations and a negative control with no protein, were prepared in either pH 4 or pH 10 buffers, and were drop cast and incubated for 2 hours on PSi films with three different pore sizes, resulting in six combinations of pore size and pH in the sensing array. The reflectance spectra were measured before and after protein solution exposure, and spectral shifts indicative of infiltration and adsorption in the pores were transduced by processing the spectra using Morlet wavelet phase analysis.[16] From the response of each sensor in the array (Figure 2), it is clear that the differences between proteins is subtle at low concentrations, but easily discriminable by eye at the highest concentrations, allowing several observations to be made. Firstly, the response to all protein solutions was proportional to pore size: as the pore size distribution shifted to higher diameters and the average pore size increased, the response increased, as expected. Notably, the relationship between response and pore size for a given protein is not linear, and there are three regimes to consider: (1) at higher average pore sizes, proteins experience essentially uninhibited entry and diffusion into the majority of the pores, (2) at lower average pore sizes comparable to the protein size, inhibited molecular transport begins to pinch off the response because there are few pores large enough to permit infiltration and adsorption, and (3) at intermediate pore sizes, there is a transition between the other two regimes.

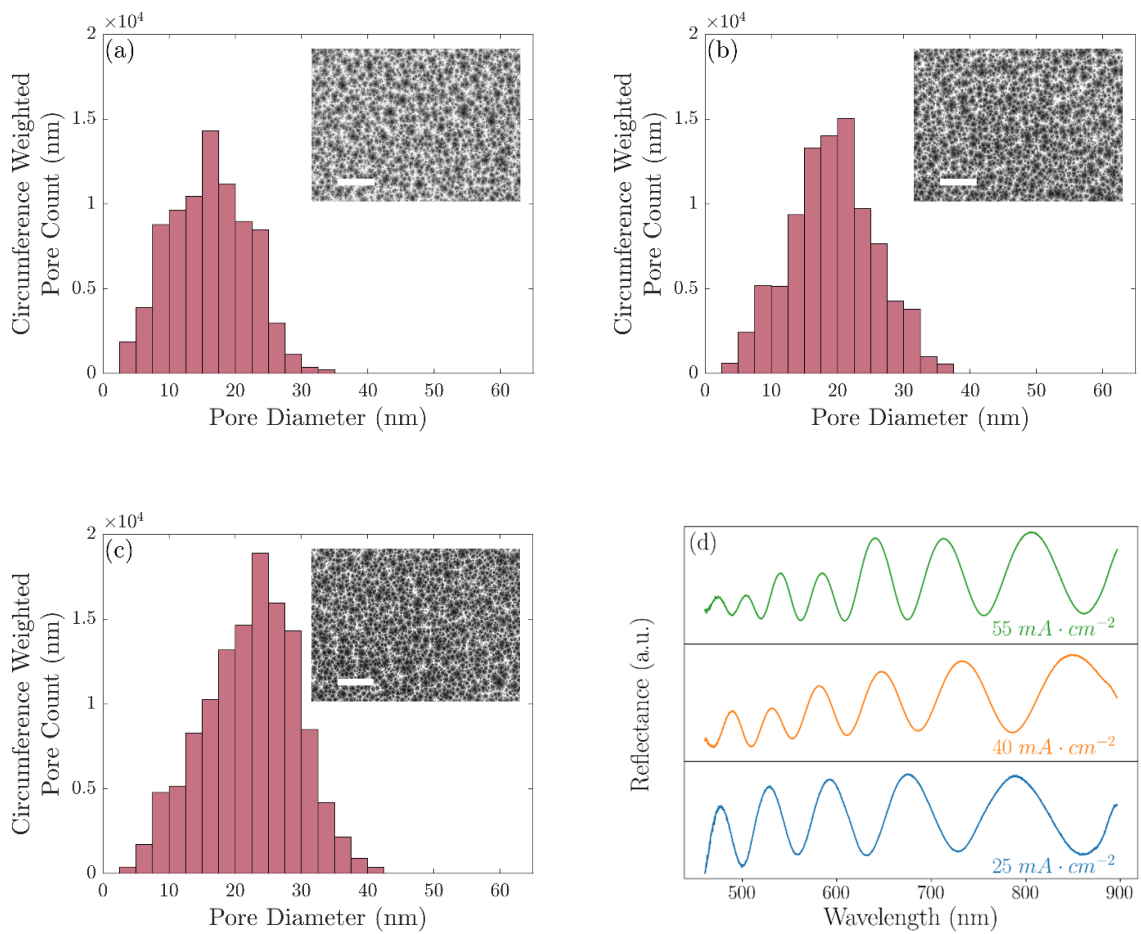

**Figure 1.** Pore diameter distributions and SEM top view images for PSi films formed with current densities of (a) 25 mA cm$^{-2}$, (b) 40 mA cm$^{-2}$, and (c) 55 mA cm$^{-2}$. (d) Measured reflectance spectra for each of these PSi films. Scale bars on SEM images are 500 nm.

**Table 1.** Measurements of average pore size, porosity, thickness, and fraction of pores larger than 30 nm for each of the PSi thin films in the sensor array, obtained through image analysis of cross sectional and top view SEM images and optical reflectance measurements.

| Etching Current Density | 25mA cm$^{-2}$ | 40mA cm$^{-2}$ | 55mA cm$^{-2}$ |
|---|---|---|---|
| Mean Pore Diameter (nm) | 12.0 | 15.1 | 17.3 |
| Thickness (μm) | 1.78 ± 0.01 | 1.94 ± 0.01 | 2.04 ± 0.01 |
| % Porosity | 53.0 | 61.3 | 66.4 |
| Fraction of Pores > 30 nm | 0.2% | 2.0% | 5.8% |

A second observation can be made regarding the effect of pH on the response of the PSi films. Multiple studies have shown that the maximum infiltration of proteins in the pores occurs when the pH environment is at the isoelectric point of the protein, resulting in a net neutral molecular charge. This condition provides minimum inhibition of protein transport and promotes close packing in the pores by avoiding extensive protein-protein and protein-PSi interactions.[15,17] Accordingly, the results in Figure 2 show that the largest response of the PSi to each of the proteins occurs when the pH of the solution is approximately equal to the isoelectric point of that given protein. We note that the surface of oxidized PSi is negative when the pH of the environment is above 2, which is the case for all experiments carried out in this work.[18] We further note that while the pH at which maximum adsorption occurs is indicated simply by a molecule's isoelectric point, the dependence of adsorption characteristics on pH will generally have a different shape for every molecule, providing another fingerprinting mechanism: properties governing infiltration and adsorption such as protein charge distribution,[19] agglomeration, and conformational changes[20] are unique to any given molecule and are pH dependent.

Thirdly, by combining the first two observations, we can understand that at pH 4, which is close to the isoelectric point of both chicken ovalbumin and BSA, the higher response is given by chicken ovalbumin due to its smaller molecular size. Finally, for solutions using pH 10 buffer, the baseline is negative due to oxidation and dissolution of the PSi matrix by the hydroxide ions present in the basic protein solutions,[21] which competes with the rising response due to protein adsorption in the pores.

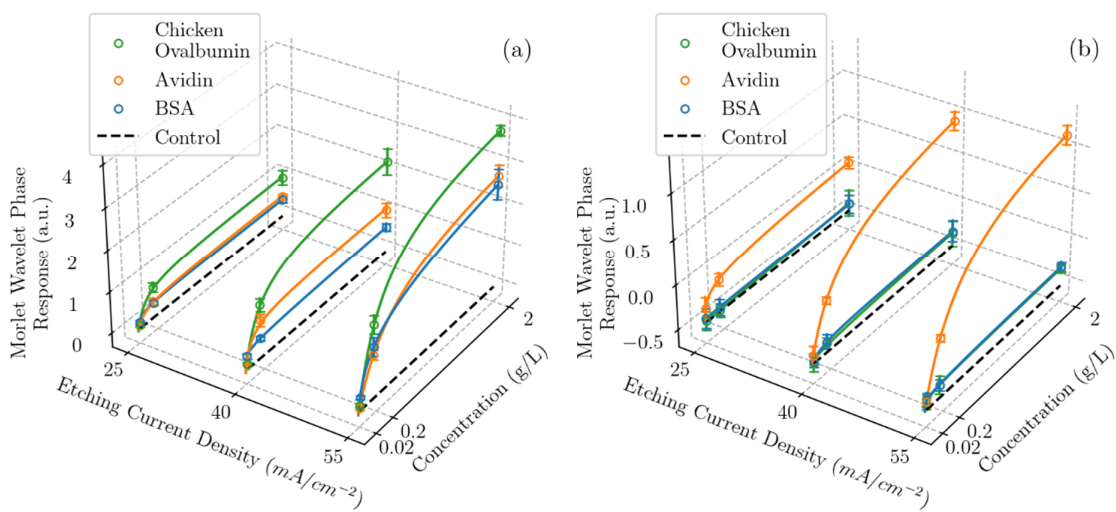

**Figure 2.** Morlet wavelet phase response as a function of both average pore size and concentration for three proteins – chicken ovalbumin, BSA and avidin – and a negative control with no protein, in solutions of DI water and (a) pH 4 and (b) pH 10 buffer, in a ratio of 0.2:0.8 (v/v). The data points represent the average value of sixteen measurements taken at the same condition and the error bars represent the standard deviation of the measurements. Each response curve was fit with the Redlich-Peterson adsorption isotherm.

Following the optical measurements and Morlet wavelet phase analysis, LDA was used to reduce dimensionality of the sensor array response matrices from six-dimensional (due to the combination of three unique formation conditions of the PSi films and two pH values used in the experiments) to three-dimensional, enabling better visualization of the degree to which the three proteins can be separated. LDA is a statistical method that determines a series of linear projections of a given training dataset that best separate data points by their associated class. This is done by maximizing the ratio of between-class variance to within-class variance. Figure 3 shows the PSi sensing array response to different concentrations of the target proteins, projected along the dominant 3 canonical factors given by LDA with 95% confidence ellipsoids overlaid. Generally, the magnitude of the variance of repeated measurements, which determines the size of the clusters and confidence ellipsoids, arises from the nature of the adsorption phenomenon rather than inconsistencies in the PSi films, reflectance measurements or experimental conditions. We note that one assumption of LDA is homoscedasticity, which is not strictly valid for this dataset: covariance is not homogenous across proteins or concentrations. For example, responses to repeat experiments using high concentration protein solutions have much higher variance than for low concentrations. Despite the violation of this assumption, LDA is still very effective at giving a visual representation of the separation between proteins and different concentrations (Figure 3). Each discriminant, or canonical factor, represents a proportion (expressed as a percentage) of the total ratio of between-class to within-class variance, which we refer to herein as discriminative power.

The first, second and third canonical factors given by LDA account for 68.7%, 28.0% and 3.1% of the discriminative power, respectively. As expected, to represent the majority of the discriminative power (eg. 95%), two canonical factors are needed, indicating that there are at least two independent properties with a significant effect on sensor response, which can be independently probed by tuning pore size and buffer pH. This simple array therefore provides more discriminatory potential than many physical sensing arrays. An additional larger pore size (obtained using an etching current density of 70 mA cm$^{-2}$) was included in preliminary experiments but was found to provide negligible additional discriminatory value due to high correlation with responses to neighboring pore size features, namely PSi films fabricated using a current density of 55 mA cm$^{-2}$.

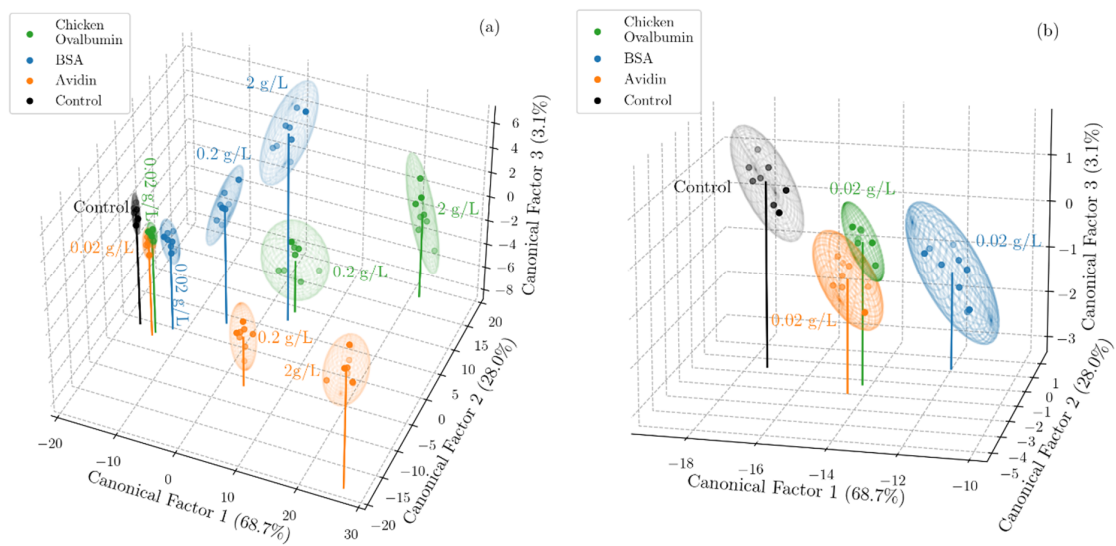

**Figure 3.** Canonical score plot of the three dominant factors obtained from LDA for (a) 3 proteins (chicken ovalbumin, BSA and avidin) at 3 concentrations (2g/L, 0.2g/L, 0.02g/L) and a negative control with no protein, and (b) the same 3 proteins at the lowest concentration (0.02g/L) and a negative control. The ease of classification and quantification of the proteins at the higher concentrations can be observed, as well as the separability at low concentration.

**Table 2.** Weightings of the original 6 dimensions corresponding to the six unique experimental conditions used to project the original high dimensional data onto each canonical factor.

|  | Canonical Factor 1 (68.7%) | Canonical Factor 2 (28.0%) | Canonical factor 3 (3.1%) |
| --- | --- | --- | --- |
| 55mA cm$^{-2}$, pH4 | 0.76 | 0.63 | 0.70 |
| 55mA cm$^{-2}$, pH10 | 0.43 | -0.88 | -0.08 |
| 40mA cm$^{-2}$, pH4 | 0.39 | 0.17 | -0.54 |
| 40mA cm$^{-2}$, pH10 | 0.25 | -0.32 | 0.01 |
| 25mA cm$^{-2}$, pH4 | 0.18 | 0.18 | -2.15 |
| 25mA cm$^{-2}$, pH10 | 0.05 | -0.07 | -0.57 |

Table 2 shows the weightings of each of the features of the original 6D response matrices which correspond to each canonical factor. For example, the response of the sensor element etched using 55mA cm-2 and exposed to protein solution in pH4 buffer is the highest contributing feature to the first canonical factor, whereas the sensor element etched with 25mA cm-2 and using pH10 buffer has almost negligible contribution to that canonical factor. These weightings can be interpreted with reference to the LDA score plot (Figure 3). From Figure 3a we can observe that this first canonical factor predominantly separates out different concentrations of the proteins rather than separating out the different types of proteins themselves. Returning our attention to Table 2, we can then understand why the largest weighting in the first canonical factor is the feature representing the largest average pore size at pH 4. These are the conditions that maximize response across all proteins, by enhancing molecular transport based on molecular weight and isoelectric point, and therefore give the largest differential between high and low concentrations. Visually, from Figure 3a, the second canonical factor separates avidin from the other two proteins and hence for this canonical factor, the weighting with the largest magnitude is the largest average pore size at pH 10, generating the largest differential between different isoelectric points. Furthermore, the weightings have opposite signs for pH 4 and pH 10. Consequently, the projection of avidin onto canonical factor 2 will be negative, given that it has a much higher relative response at pH 10, whereas the projection of BSA and chicken ovalbumin onto canonical factor 2 will be positive due to their relatively high response in pH 4 conditions. The third canonical factor primarily separates molecules by molecular weight according to Figure 3a. The sign of the weighting enhances the differential response since the higher molecular weight molecules (BSA and avidin) have a relatively high response to the higher average pore size sensors, but a much lower relative response at lower average pore size sensors for which the weighting is negative. On the other hand, chicken ovalbumin has a much higher relative response at the lower average pore size sensors as well as the high pore size. Consequently, BSA and avidin responses will mostly lie above zero in the 3rd canonical factor axis, whereas the response to chicken ovalbumin will largely reside below zero.

Next, the concentrations of each of the three proteins were classified with support vector machines (SVMs) using a linear kernel and regularization hyperparameter C = 100 (informed by analysis of a small preliminary dataset). The SVMs were trained on the reduced dimensionality dataset given by LDA, which improved accuracy by reducing noise. SVMs are supervised optimal margin classifiers, chosen due to their stability, interpretability, and applicability to small datasets. The model, when coupled with leave-one-out cross validation, gave accurate predictions of both protein type and concentration in 100% of the cases when results from two sensing arrays were averaged. To test the ability of the model to classify unseen concentrations, the same SVM model with a linear kernel was retrained to classify protein type only, and a test set was compiled by carrying out further experiments in which sixteen sensor arrays were exposed to chicken ovalbumin at a concentration of 0.1 g/L. By averaging two measurements, these further experiments yielded eight new data points. This choice of protein and concentration was made to rigorously test the system. It is clear that avidin can be trivially classified due to the large differential effect of pH, whereas, in the training dataset (Figure 3), chicken ovalbumin and BSA are less easily discriminable, particularly at low concentrations. The independent test set was classified with an 87.5% accuracy (one protein misclassified of the eight in the test set), illustrating thepromise of this approach for classifying proteins of unknown concentration. We note that to avoid data leakage, the mean and standard deviation of the training data set were used to standardize all data, both test and training sets. For the same reason, the canonical factors given by LDA analysis when applied to solely the training set were used to transform the test set. The same process was followed in the case of leave-one-out cross validation for which the data point left out of the training set for classification is considered the test set. The accuracy of classification using these SVM models could be increased with a larger array incorporating more pore sizes and pH values to give more discriminating power. Additionally, because no cross validation for model selection or hyperparameter tuning was carried out due to the limited amount of data, accuracies reported here are a lower bound of what could be achieved with more data and a more complex optimized model. We note that while the 3rd canonical factor in Fig. 3 represents a small percentage of the discriminatory power (3.1%), it plays a critical role in separating BSA and chicken ovalbumin which have similar pI but different molecular weights, and almost identical trajectories when projected only on to the first 2 canonical factors. Consequently, the third canonical factor will have a large contribution to the accuracy of models trained on the dimensionality reduced training set. As a result, the discriminatory power is not the best indicator of feature importance in the context of discriminating proteins, partly a consequence of investigating different concentrations of each target molecule.

For more complicated sensing scenarios, more discriminatory and quantitative power can be realized by increasing the size of the sensing array, encompassing additional degrees of freedom to distinguish between a larger number of proteins and other species of interest. This could be achieved by widening the parameter space to include ionic strength of solution, surface charge and hydrophobicity, in addition to using more complex PSi structures and monitoring the adsorption and diffusion in real-time, which all serve to increase dimensionality of the dataset. Existing sensor arrays differentiate between molecules predominantly through chemical interactions with surface coatings: PSi can support a diverse range of robust surface chemistries so the sensor design reported here can be expanded to leverage this additional parameter space, giving rise to a sensing array which incorporates both chemical and physical interactions, and has discriminatory potential above and beyond what is possible when relying solely on surface coatings. Furthermore, the cost effective and scalable nature of PSi sensor arrays is a key advantage, and large sensor arrays feasibly could be measured with an imaging system such as a smartphone camera, enabling the response of an almost arbitrarily large array of PSi sensors to be captured as a function of time. Finally, we note that our capture agent-free sensing approach is not intrinsically tied to PSi but can be applied to a wide range of porous material platforms, including porous anodic alumina[22] which offers a narrower pore size distribution within a given film.

## 3. Conclusion

We report a sensing system based on an array of porous silicon sensors, each with a unique combination of properties but no functionalization or capture agents. This system is able to classify and quantify three proteins with similar molecular size down to concentrations of ~300nM using LDA for dimensionality reduction and SVMs for prediction and classification. An accuracy of 100% was achieved for proteins and concentrations previously encountered in the training set, and a previously unseen independent test set collected using an intermediate concentration of one of the proteins was classified with 87.5% accuracy. The design of this system could obviate the need for capture agents, paving the way for cheaper, more robust and quicker to develop sensors that provide medical diagnostics, environmental monitoring and food safety systems to resource limited environments.

## 4. Methods

*Materials*: Chemicals were all analytical grade and used without further purification. De-ionized (DI) water (resistivity 15 MΩ cm), used for all solutions, was produced using a Millipore Elix water purification system. Single side polished, boron doped silicon wafers (⟨100⟩, 0.01−0.02 Ω cm, 500−550 μm) were purchased from Pure Wafer. Ethanol, BSA, chicken ovalbumin, and avidin were purchased from Thermo Fisher Scientific, and pH 4 and pH 10 reference standard buffers were obtained from Sigma-Aldrich. Aqueous hydrofluoric acid (HF) (48-51%) was purchased from Acros Organics. The pH of all solutions was measured using a Mettler Toledo Seven Easy pH-meter.

*Preparation of Single Layer PSi*: Single layer PSi thin films were fabricated by electrochemically etching p-type silicon wafers using HF, described in detail elsewhere.[21] A solution of HF in ethanol (15% v/v) was used to etch the wafers in an Advanced Micromachining Tools (AMMT) MPSB PSi wafer etching system. Firstly, a sacrificial layer was etched using a current density of 70 mA cm$^{-2}$ for 100 s, which was subsequently dissolved in 1 M KOH solution. Secondly, the wafer was washed with DI water and ethanol to remove HF residue and then etched again at a current density of either 55 mA cm$^{-2}$, 40 mA cm$^{-2}$, or 25 mA cm$^{-2}$, to form thin films with different pore size distributions for the different elements in the sensing array. The etching time used to fabricate the PSi films (57s, 66s and 93s for 55 mA cm$^{-2}$, 40 mA cm$^{-2}$, and 25 mA cm$^{-2}$, respectively) was tailored to give approximately the same thickness regardless of etching current density (and associated unique etch rate). Thirdly, the wafer was diced into square 5 × 5mm samples using a DISCO DAD3220 dicing saw. Finally, the samples were oxidized at 800°C in ambient air for 10 min, forming a passivating surface layer of $SiO_2$, which is hydrophilic, and accumulates a negative surface charge in pH > 2 conditions.

*Material Characterization:* Properties of the PSi films were measured by analyzing scanning electron microscope (SEM) top view and cross sectional images, including the pore size distribution and mean pore size, porosity and thickness (Table 1) and pore size distribution (Figure 1). To extract pore distribution and average pore size, analysis was carried out in MATLAB.[23] First, the contrast of the top view SEM images was made uniform across the image and enhanced using the adapthisteq MATLAB function,[24] and a threshold is used for conversion to a binary image. Isolated pixels were removed and a median filter was applied. The perimeter and area of each of the pores was found using the regionprops MATLAB function. The pixel to nm conversion is performed using the scale bar in the SEM images. The count of pores in each bin of the pore distribution histogram is weighted by the average perimeter length for each of the pores in that bin.

*Optical Reflectance Measurements*: Reflectance spectra were measured by coupling light from quartz tungsten light source into a bifurcated optical fiber through one fiber port and measuring the reflected light using an Ocean Optics USB 4000 CCD spectrometer connected to the second fiber port. The height of the fiber was adjusted to form a spot size 5 mm in diameter on the PSi sensor surface. The output spectra from the spectrometer were fed to a PC running the Ocean Optics Spectra Suite software15, which averaged 100 spectra and saved the result once per second.

*Experimental Procedure*: PSi samples (5 × 5 mm) were washed with water and ethanol, and dried under nitrogen. Reflectance spectra of the six sensing elements in an array were measured before protein incubation to establish a baseline reference spectrum. Three concentrations of each protein solution were prepared (2 g/L, 0.2 g/L 0.02 g/L) in 20% DI water 80% reference standard buffer (pH 4 or pH 10) solutions, which corresponds to concentrations of 30 µM, 3 µM and 300 nM for BSA (pI = 4.63, MW = 66.4 kDa), 45 µM, 4.5 µM and 450 nM for chicken ovalbumin (pI = 4.54, MW = 44.3 kDa) and 30 µM, 3 µM and 300 nM for avidin (pI = 10, MW = 66-67 kDa). A 20 µL volume of protein solution was drop cast on each sensing element in the array, and left to incubate for 2 hours, followed by a wash in a water bath for 10s. The purpose of the wash is to remove unbound molecules from the PSi surface and inside the pores. A much smaller number of weakly bound molecules will also be removed, but the reflectance change during washing is small, and almost entirely independent of wash duration, indicating that most molecules are adsorbed strongly enough to remain in the pores. Each sensing element in the array was then dried under nitrogen and measured again, and the resulting spectrum was compared to the reference spectrum before protein solution exposure by calculating the Morlet wavelet phase response.[16]

**Acknowledgements**

The MATLAB code used to measure pore sizes and calculate pore size distributions from greyscale images is freely downloadable from the Weiss group website (https://my.vanderbilt.edu/vuphotonics/resources) alongside a readme file with user instructions. The authors acknowledge the Vanderbilt Institue for Nanoscience and Enginnering (VINSE) for providing facilities for PSi etching, wafer dicing, and SEM imaging.